\documentclass[]{emulateapj}

\shorttitle{Microlensing of Circumstellar Disks}
\shortauthors{Z. Zheng \& B. M\'enard}

\begin{document}
\title{Microlensing of Circumstellar Disks}
\author{Zheng Zheng\altaffilmark{1} and Brice M\'enard}
\affil{Institute for Advanced Study, Einstein Drive, Princeton, NJ 08540, USA}
\altaffiltext{1}{Hubble Fellow}

\def \etal{et~al.\/}
\def\RE{\hat r_{\rm E}}
\def\tE{t_{\rm E}}

\begin{abstract}
We investigate the microlensing effects on a source star surrounded by
a circumstellar disk, as a function of wavelength. The microlensing
light curve of the system encodes the geometry and surface brightness
profile of the disk. In the mid- and far-infrared, the emission of the
system is dominated by the thermal emission from the cold dusty disk.
For a system located at the Galactic center, we find typical magnifications 
to be of order 10-20\% or higher, depending on the disk surface
brightness profile, and the event lasts over one year.  At around 20
$\mu$m, where the emission for the star and the disk are comparable,
the difference in the emission areas results in a chromatic
microlensing event.  Finally, in the near-infrared and visible, where
the emission of the star dominates, the fraction of star light
directly reflected by the disk slightly modifies the light curve of
the system which is no longer that of a point source.  In each case,
the corresponding light curve can be used to probe some of the disk
properties.  A fraction of $10^{-3}$ to $10^{-2}$ optical microlensing
events are expected to be associated with circumstellar disk systems.
We show that the lensing signal of the disk can be detected with
sparse follow-up observations of the next generation space telescopes.
While direct imaging studies of circumstellar disks are limited to the
solar neighborhood, this microlensing technique can probe very distant
disk systems living in various environments and has the potential to
reveal a larger diversity of circumstellar disks.
\end{abstract}


\section{Introduction}

During the last decade, gravitational microlensing has become a
powerful tool for analyzing the distribution and nature of compact
dark objects in the halos of our and nearby galaxies. The idea was
first suggested by \citet{paczynski1986} and nowadays a number of
monitoring programs are ongoing and have detected several thousands of
microlensing event candidates (\citealt{alcock1993,alcock1996,aubourg1993,
udalski1993,udalski1994,udalski2003,tisserand2005,afonso2003a,afonso2003b,
alcock2000a,alcock2000b,popowski2003,calchi-novati2005}). Microlensing can 
provide us with interesting constraints on the mass of the lenses (e.g., 
\citealt{an2002,gould2004,jiang2004,kubas2005}) and the presence of
their planetary companions (e.g., \citealt{mao1991,gould1992,gaudi2002,
albrow2001c,bond2004,abe2004,udalski2005}) and circumstellar structures
(\citealt{Bozza02a,Bozza02b}). In addition, several authors 
have pointed out that microlensing events themselves can yield useful 
information about the {\it sources}, i.e., the lensed stars, by taking into 
account their angular extent (e.g., \citealt{gould1994,simmons1995,
sasselov1997,valls-gabaud1998,hendry1998,heyrovsky1997,
Gaudi1999,Cassan2004,albrow2001b,Castro2001,Gould2001}).
For example, the 
limb-darkening effect of the source star can be used to constrain models 
of stellar atmosphere, and has been measured in a few microlensing events 
(e.g., \citealt{afonso2000,albrow2000,albrow2001a,fields2003}). 

In this paper, we discuss another microlensing application: the 
magnification of a star surrounded by a circumstellar disk. The
microlensing properties of such a system appear to be quite
interesting: in the optical and near infrared bands, the emission is
dominated by the star light and classical \citeauthor{paczynski1986} light curves can be
observed, whereas in the mid/far-infrared, the emission from the cold 
dust of the disk dominates and produces more complex light curves.
Microlensing translates spatial structures into temporal ones and allows us to ``resolve'' the disk.
Combining information from both wavelength ranges can place constraints 
on some of the disk properties. In addition, because of the presence of 
a disk, a fraction of the star light is reflected by the dusty particles. 
As this light contribution no longer originates from a point-like source, 
it will cause the microlensing light curve to slightly deviate from that 
for star-only in the optical and near-infrared. This effect can also be 
used to constrain disk properties. Direct imaging studies of disks are 
possible but are limited, in the foreseeable future, to the vicinity of 
the solar system. The microlensing technique allows us to access systems 
that are much further away, i.e., up to $\sim$ 10 kpc. This technique can 
probe circumstellar disks in various environments with different 
metallicities and has the potential to reveal the diversity of such systems.

The characteristics of circumstellar disk microlensing light curves
differ from those of stars: the total magnification of the disk in the
mid-infrared is expected to be significantly smaller than that of a
star in the optical. In addition the relative duration of the
microlensing event for the disk is expected to be much longer.
Indeed, the typical radius of a circumstellar disk $a$ is of the order
of a few tens of AU, though it depends on the age and type of the
central star \citep{zucker01}. For a source at the Galactic center and a
lens of 1 M$_\odot$ located halfway, the Einstein ring radius $\RE$ in the
source plane is about 8AU, which is much smaller than the source (disk)
size. We therefore have an extreme finite source effect and only that
part of the source inside the Einstein ring is efficiently magnified. For
a large uniform face-on disk, the expected overall fractional flux excess 
caused by the lensing effect is $2(\RE/a)^2$ \citep{agol2003}, which is at 
the level of $\sim 10\%$ if the disk radius is a few times larger than 
$\RE$. The microlensing timescales are also different. For a
star, the duration of the lensing event $\tE$, i.e., the time required to
cross the Einstein radius, is about one month.  For a circumstellar
disk, this timescale is about $(a/\RE)\tE$, which ranges between half
a year to a couple of years. If a microlensing event of a (young) star
is discovered (in optical or infrared), one can then (sparsely)
monitor the object in the mid-infrared and look for the excess
characteristic of circumstellar disks. If a disk can be detected, then
observing its flux as a function of time will provide us with
constraints on the disk size and brightness distribution as the lens
differentially resolves the disk. 

In this paper we show that next generation space telescopes will be
able to detect microlensing events of circumstellar disks at the Galactic 
center.  The outline of this paper is as follows. In \S~2, we present 
some characteristic parameters of circumstellar disks. In \S~3, we
introduce the formalism describing the observed light curve of a disk. 
In \S~4, we show how the disk parameters are related to the shape of 
the light curve and the behaviors of light curves in different 
observational bands. We then investigate the detectability and the 
occurrence of this phenomenon in \S~5. Finally, we summarize and discuss 
our results in \S~6.

\section{Characteristics of Circumstellar Disks}

Circumstellar disks evolve through two main stages: during the first
phase, the protoplanetary disk composed of gas and dust is optically
thick and provides a particle reservoir for planet formation. By the
time stars of $\sim1\,\rm M_\odot$ reach ages of $\sim1\,$Myr, the
accretion disk surrounding the stars has lost its gas, becomes
optically thin and readily accessible to observation at many
wavelengths.  It is referred to as a debris disk. The dust particles
in the disk absorb a fraction of the star's photons and re-emit them
at longer wavelengths. The parameters of some debris disks observed in
the vicinity of the solar system are listed in \citet{zuckerman2004}.
Current observations show that typical systems have a star temperature
of about 6000 K and the excess infrared emission from the disk can
be well fitted by a blackbody spectrum with an effective temperature of 
about 80 K. This does not imply that the disk is isothermal but is simply 
a useful way for describing the disk spectral energy distribution.

Apart from the re-emitted light, some of the star photons intercepted by 
the disk are reflected. The relative amount of light being reflected and 
that being absorbed and re-emitted depends on the albedo of the disk,
which is of the order of a few tens of percent. For simplicity, we assume 
the albedo to be 0.5 so that the disk seen through the reflected light and 
through the re-emitted light has the same luminosity, which we denote 
as a fraction $\tau$ of the star's bolometric luminosity.  

\begin{figure}
\plotone{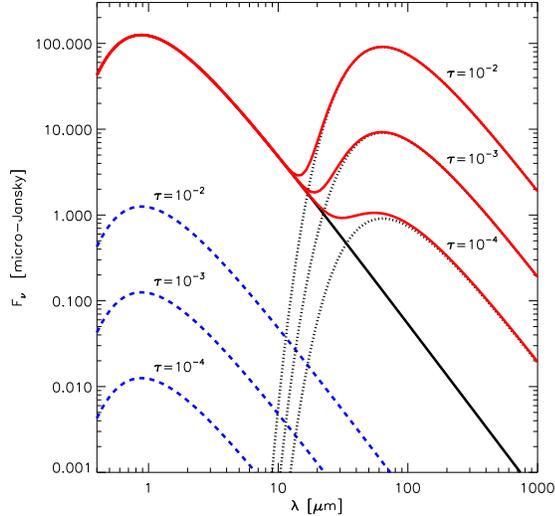}
\caption[]{\label{fig:sed} 
Spectral energy distribution of a star with $T=5800 K$ surrounded by a 
dusty debris disk with $T=80K$. This system is put at the Galactic center 
(distance $D=8$ kpc). The reflected (dashed curves) and re-emitted (dotted 
curves) lights from the disk are calculated for three values of the 
disk-to-star luminosity ratio: $\tau=10^{-2}$, 10$^{-3}$, and 10$^{-4}$.}
\end{figure}

It has been empirically found that the fraction $\tau$ depends on the age 
of the system, $\tau\propto {\rm age}^{-1.75}$ for ages between 
$10^{6.5}$ and $10^{9.5}$ years \citep{zuckerman2004}. For a 10 Myr system
and a 100 Myr system, $\tau\sim10^{-2}$ and $\sim 2\times10^{-4}$, 
respectively. The expected spectral energy distribution of such systems 
located at the Galactic center (a distance of $D=8$ kpc), is illustrated in 
Figure \ref{fig:sed} for different values of $\tau$. 
In this plot we have not considered extinction effects as they are in general
small for the wavelength ranges we consider below.
As can be seen, the 
infrared excess caused by the thermal emission of the disk decreases rapidly 
as the system ages. In the figure we have assumed that the reflected light 
from the disk has the same spectral energy distribution as the star. 

Circumstellar disks are expected to have an inner disk radius set by
the sublimation temperature of the dust grain \citep{natta2001} with radii ranging from around
one to a few AU, and in addition, theoretical models predict that
planets in disks can interact with the disk material and open wide
radial gaps (\citealt{pap84,paar04}), once they achieve sufficient
mass. The width of the gap is expected to be $\sim$20\% of the orbital
radius, which would decrease the disk surface density in a 1 AU
annulus at 5 AU for example. 
In addition disk structures such as rings, warps and blobs are
expected as a result of the presence of planets (\citealt{Moran2004,
Kuchner2003,Wilner2002,Wyatt2002,Quillen2002,Wahhaj2003,Koerner2001})
and will be imprinted in the microlensing light curve of the system
and, in some cases, might enhance the detectability of the disk.

At the Galactic center, the angular size of a 30 AU disk is only $\sim
0.004''$. Spatially resolving such system in the visible and
mid-infrared would require a telescope larger than 30 m, and 1 km
respectively. With microlensing, however, the spatial information is
translated into a temporal one. As shown below, measuring the
microlensing light curve allows us to probe the disk geometry and 
surface brightness profile.

\section{The formalism of microlensing of disks}
\label{sec:basics}

The characteristic scale for microlensing is the Einstein radius of
the lens. For our purpose in this paper, the Einstein radius is defined
in the {\it source} plane, which reads
\begin{eqnarray}
\RE &=& \left [ \frac{4 G M}{c^2} \frac{(D_s - D_l) D_s}{D_l} \right ]
        ^{1/2}\nonumber\\
    &=& 8.1\, {\rm AU} \left(\frac{M}{M_\odot}\right)^{1/2} 
        \left(\frac{D_s}{8 {\rm kpc}}\right)^{1/2} 
        \left(\frac{1-x}{x}\right)^{1/2},
\end{eqnarray}
where $M$ is the lens mass, $x=D_l/D_s$, and $D_s$ and $D_l$ are the
distances to the source and lens, respectively.  For a source at the
Galactic center and a 1 M$_\odot$ lens located half way, $x=1/2$ and
$\RE\simeq8$ AU. The corresponding time scale for the source to cross the 
Einstein radius is
\begin{eqnarray}
\label{eqn:t_E}
\tE & = & \RE/v  \nonumber \\
    & \simeq & 70\, {\rm days} \left(\frac{M}{M_\odot}\right)^{1/2} 
        \left(\frac{D_s}{8 {\rm kpc}}\right)^{1/2} \nonumber\\
    & & \times \left(\frac{1-x}{x}\right)^{1/2}
\left(\frac{v}{200 {\rm km/s}}\right)^{-1},
\end{eqnarray}
where $v=D_s\mu_{\rm rel}$ and $\mu_{\rm rel}$ is the relative proper motion 
between the source and the lens.

Considering the central star as a point source, the
magnification $A$ of the source can be shown to depend only on the
projected separation $u$ of the lens and source, in units of $\RE$ of
the lens \citep{paczynski1986}, namely
\begin{equation}
A(u) = \frac{u^2 + 2}{u\,\sqrt{u^2 + 4}}.
	\label{eq:amp}
\end{equation}
The relative motion between the source and the lens changes $u$ and
therefore the magnification varies as a function of time. Microlensing 
events of point sources can be identified from this characteristic light
curve.

\citet{heyrovsky1997} considered the microlensing of inclined accretion
disks of active galactic nuclei. In this case, the Einstein ring radius is 
much larger than the size of the source and large magnification values can 
be reached. The case considered in this paper is different: the size of 
Galactic circumstellar disks is larger than the Einstein ring radius leading 
to lower magnifications and longer timescales.  For the magnification of 
very large sources, a number of analytical approximations have been derived 
in the literature (\citealt{heyrovsky1997}, \citealt{agol2003}).
In our case, the Einstein ring is not necessarily much smaller
than the spatial extents of features in the disk, therefore we proceed with
the exact calculations.

\begin{figure}[h]
\plotone{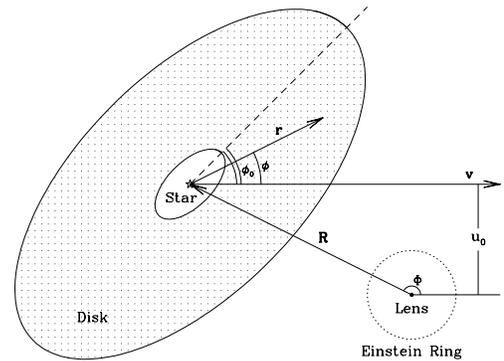}
\caption[]{\label{fig:geometry}
Geometry of lensing of a circumstellar disk. The dotted ring around the lens
indicates the Einstein ring. The circumstellar disk is assume to have an hole
inside an inner radius and a cutoff outside an outer radius. Projected onto 
the sky plane, the disk appears to be an ellipse with the orientation angle 
of the major axis being $\phi_0$. Any point on the disk is specified by its
coordinates $(r,\phi)$ in a polar system centered at the star. The angles 
$\phi_0$ and $\phi$ are with respect to the direction of the relative 
motion ${\mathbf v}$. The position of the star (or the disk center) is 
specified by $(R,\Phi)$ with respect to the lens ($\Phi$ is relative to the
direction of ${\mathbf v}$). The impact parameter of the star is denoted as 
$u_0$.
}
\end{figure}

In this paper we do not intend to model the disk in great details.
For estimating the lensing signal we will use different surface brightness 
profiles in order to represent the changes in dust and temperature
distributions. Let us now consider an axially symmetric disk with
surface brightness profile $f(a)$, where $a$ is the radial distance to
the central star. The projection of the disk onto the sky plane is an
ellipse.  The configuration of the disk is fixed given its center
position $(R,\Phi)$ with respect to the lens, the inclination angle
$i$, and the orientation angle $\phi_0$ of the projected major axis
(with respect to the direction of the relative motion). The geometry
and meanings of symbols are shown in Figure~\ref{fig:geometry}.  In
all our discussions below, quantities with length dimension are in
units of the Einstein ring radius $\RE$ and surface brightnesses are
in units of flux/$\RE^2$.  We consider a geometrically thin disk with
isotropic emission at every point. With no lensing effect, the total
flux of the disk can be calculated without projection and is given by
\begin{equation}
 F_0 = \int\int f(a) a da d\phi = 2\pi\int f(a) a da.
\end{equation}
If the disk is lensed, the total flux is the unlensed flux convolved with
the magnification profile $A(u)$,
\begin{equation}
\label{eqn:F}
F =\int\int A(u) f_p(r,\phi) r dr d\phi,
\end{equation}
where $f_p$ is the projected surface brightness profile,
$u=[R^2+2Rr\cos(\Phi-\phi)+r^2]^{1/2}$ is the distance from the lens
to a (projected) disk element. Here $(r,\phi)$ are polar coordinates
with origin at the center of the star (see
Figure~\ref{fig:geometry}). The integration becomes easier if we
change the variable $r$ to the radius $a$ along the major axis
according to the relation
$a=r\sqrt{1+\sin^2(\phi-\phi_0)\tan^2i}$. The projected surface
brightness $f_p$ is simply $f(a)/\cos i$. Equation~(\ref{eqn:F}) then
has the form
\begin{equation}
\label{eqn:convolution}
 F = \int\int A(u) \frac{f(a)}{\cos i} \frac{a}{1+\sin^2(\phi-\phi_0)\tan^2i}
 da d\phi.
\end{equation}
The overall magnification of the disk flux is $F/F_0$. The temporal dependence 
of the magnification is through the position of the disk center $(R,\Phi)$ 
with respect to the lens,
\begin{equation}
R(t)=\sqrt{t^2+u_0^2}~~\mbox{  and  }~~
\Phi(t) = \arccos \frac{t}{\sqrt{t^2+u_0^2}},
\end{equation}
where $u_0$ is the impact parameter in unit of $\RE$, and $t$ is the time in 
unit of $\tE=\RE/v$ defined in equation~(\ref{eqn:t_E}). We set $t=0$ at 
$\Phi=\pi/2$.

From equation (\ref{eqn:convolution}), we see that the light curve of the 
disk depends on the magnification kernel $A(u)$ and the geometry and the 
surface brightness profile of the disk. It should be noted that 
$A(u)$ can be directly inferred from observations of the light curve
of the star in optical or near-infrared band.  The disk light curve
can be used to constrain $f(a)$ and the geometric parameters only.

Before investigating the expected light curves described by the exact 
results of equation~(\ref{eqn:convolution}), it is interesting to get a 
more intuitive understanding of this equation by considering the following
approximation: if the spatial features of the disk are much larger than the
Einstein ring radius, the calculation of the magnification can be simplified 
by approximating $A({\mathbf u})-1$ as $2\pi\delta_D({\mathbf u})$, where 
$\delta_D$ is the Dirac-$\delta$ function. The increase in flux 
$\Delta F=F-F_0$ caused by the lensing simply traces the local surface 
brightness of the disk at the position of the lens (i.e., 
${\mathbf r}=-{\mathbf R}$). The result is
\begin{equation}
\Delta F=2\pi f_p(r,\phi) = 2\pi f(a)/\cos i,
\end{equation}
where $r=R(t)$, $\phi=\Phi(t)-\pi$, and $a$ is related to $r$ and $\phi$ as 
mentioned above equation (\ref{eqn:convolution}). The excess flux is just 
twice the Einstein ring area times the surface brightness at the lens 
position, which is the same as derived by \citet{heyrovsky1997} (see their
eq.[34]) for an elliptical source and by \citet{agol2003} (see his eq.[10])
for general large sources. In fact, the result of the magnification for
a large source presented in \citet{agol2003} (his eq.[4]) can be derived 
as above by noticing that $A({\mathbf u})-1 \simeq 2\pi\delta_D({\mathbf u})$. 
Approximate calculations of the magnification at the edge of a large source
can be found in \citet{agol2003}.

It should be noted that the light curve alone is not sufficient to determine 
the physical size of a disk and what we infer are all in units of the Einstein 
ring radius.  Additional information on the size of the Einstein ring radius
could be obtained from the star-only light curve if the finite source effect 
and the parallax effect are detected. The former gives the angular size of 
the Einstein ring radius with respect to that of the source star (e.g., 
\citealt{gould1994}) and the latter measures the Einstein ring radius 
projected onto the observer's plane (see, e.g., \citealt{gould1992a,jiang2004, 
gould2004}). Since the impact parameter needs to be
small for a measurable finite source effect, it is unlikely that the 
lensing of a system with circumstellar disk can measure the angular size of 
the Einstein ring. Even if we have both the relative angular size of the
Einstein ring and its projected size onto the observer's plane,
we still need the source distance in order to infer the size of 
the Einstein ring projected onto the {\it source} plane. The radial velocity 
of the source star could provide some information on the source distance. 
If the blended light in the microlensing event is caused by the lens, which 
can be verified astrometrically, information about the mass and distance of 
the lens can be obtained (\citealt{ghosh2004}). 
Finally, if we observe the event at different wavelengths, the temperature as a
function of radius in units of the Einstein ring radius can be
constrained. Then, by knowing the dust particle albedo, measuring the
central star temperature, and assuming thermal equilibrium, the
corresponding physical radii can be inferred. In general, we may have 
to rely on some statistical priors to constrain the physical size of the disk.

\section{Observational signatures}
\subsection{Mid- and Far-infrared: Re-emitted Light}

\begin{figure}[t]
\epsscale{1.05}
\plotone{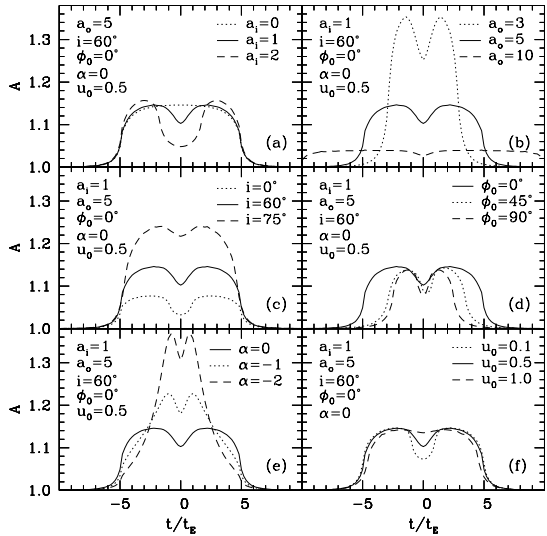}
\epsscale{1.}
\caption[]{\label{fig:lightcurve}
Lensing light curves of the disk. In each panel, one disk parameter is
varied as indicates at the upper-right corner with others fixed at values
listed at the upper-left corner. The y-axis is the magnification. 
The disk parameters include the inner radius $a_i$, the outer radius $a_o$,
the inclination angle $i$, the orientation angle $\phi_0$ of the projected
major axis, the surface brightness profile index $\alpha$, and the impact 
parameter $u_0$. Parameters with length dimension are in units of the Einstein 
ring radius.
}
\end{figure}

In the mid- and far-infrared, the luminosity of the system is dominated by 
the light re-emitted by the disk and the contribution of the star can be 
neglected. The light curve of the disk depends on its geometrical properties 
(the size, the inclination angle $i$, the orientation angle $\phi_0$ of the
projected major axis, and the impact parameter $u_0$) and its surface 
brightness profile. To show its dependence on various parameters, we
assume that the disk has an inner radius $a_i$ and an outer radius
$a_o$ with a power-law surface brightness profile $f(a)\propto
a^{\alpha}$ between these two radii. The light curves are shown in
Figure~\ref{fig:lightcurve} with $a_i$, $a_o$, $i$, $\phi_0$, $u_0$, and
$\alpha$ varied, one at a time.

As shown in panel (a), the existence of an inner radius of the disk causes a
corresponding dent in the light curve because the disk is less efficiently
magnified as the lens comes close to the inner radius. The dent becomes deeper
as the inner radius becomes larger. The width of the dent is proportional to 
the inner radius. Similarly, radial gaps opened by planets in the disk will
also exhibit such dents in the light curve. Panel (b) shows cases for which 
the outer radius varies: the smaller the disk, the larger the magnification 
as the percentage that can be effectively lensed becomes larger. The overall 
magnification is inversely proportional to $a_o^2-a_i^2\sim a_o^2$, and the 
duration time is proportional to $a_o$. 

The area of the disk projected onto the sky plane depends on the
inclination angle, i.e., proportional to $\cos i$. With other
parameters fixed, a more inclined disk would have a higher overall
magnification, as shown in panel (c), because its projected area is
smaller and it becomes more efficiently magnified.  On the other hand,
the orientation of the projected disk does not change the area, and
has therefore little effect on the maximum magnification. This is
shown in panel (d). However, the duration has a dependence on the
orientation, since it changes the path length that the lens crosses
the disk. Except for cases with $\phi_0=0^\circ$ and $90^\circ$, the
orientation of the disk (plus the non-zero impact parameter) also
causes the light curve to be asymmetric about t=0.

In panel (e), we investigate the effect of the surface brightness profile of 
the disk.  With a more concentrated surface brightness profile, the disk 
appears to be less extended, and the maximum magnification increases. In our 
example, the maximum amplitude changes from $\sim$15\% for a uniform disk to 
$\sim$40\% for a disk with surface brightness $\propto a^{-2}$. Slopes at 
the rising and fading parts of the light curve also probe the surface 
brightness profile, in the sense that a steeper light curve corresponds to 
a steeper brightness profile. As we mention in \S~\ref{sec:basics}, if the 
disk is much larger than the Einstein ring, the light curve simply traces 
the surface brightness profile along the lens trajectory.

For a point source, the magnification is determined by the impact parameter
$u_0$. However, for a disk close to uniform, the impact parameter has little 
effect on the magnification of the disk flux, since the magnification is 
largely determined by the area ratio of the Einstein ring to the projected 
disk.  Panel (f) shows that, because of the inner hole in the disk, a smaller 
impact parameter leads to a deeper dent in the light curve.

\begin{figure}[h]
\epsscale{1.05}
\plotone{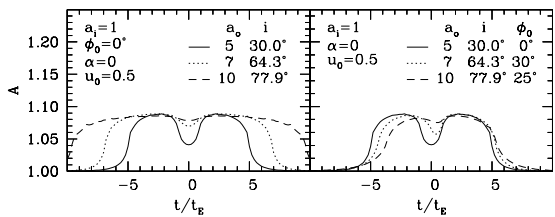}
\epsscale{1.}
\caption[]{\label{fig:lightcurve1}
Lensing light curves with similar magnification. In the left panel, values of 
the outer radius $a_o$ and the inclination angle $i$ are changed in a way to 
keep the projected area of the disk fixed [$(a_o^2-a_i^2)\cos i = constant$]. 
In the right panel, additional changes of the orientation angle $\phi_0$ are
made to match the duration time. 
}
\end{figure}

After having seen how the light curve responds to different parameter
changes, it should be noted that a number of parameter combinations
could be degenerate.  To first order, the light curve of a uniform disk
can be characterized by the maximum magnification and the duration time.
As we have discussed, the magnification is largely determined by the area 
ratio of the Einstein ring to the projected disk. For example, a larger 
disk that is highly inclined could have the same projected area as a 
smaller disk that is slightly inclined. The magnifications for the two cases 
would be similar. The degenerate direction of the magnification roughly 
follows $(a_o^2-a_i^2)\cos i = constant$. The left panel of
Figure~\ref{fig:lightcurve1} shows examples with $(a_o^2-a_i^2)\cos i$
fixed. As expected, the amplitudes of the light curves are similar. The 
duration time (in unit of $\tE$) of the event is the length (in unit of $\RE$)
of the chord the lens crosses. For a larger disk, it 
becomes longer, but this can change if we vary the orientation of the
projected disk as seen in panel (d) of Figure~\ref{fig:lightcurve}. By 
adjusting the orientation parameter $\phi_0$, we can have
light curves that mimic each other, as shown in the right panel of
Figure~\ref{fig:lightcurve1}. The asymmetry of the light curves can
break the degeneracy. But if the data are noisy, the solution would be
partially degenerate with respect to the size of the disk, the inclination
angle, and the orientation angle. 

The above degeneracy can be broken by making use of the star-only light 
curve (in optical or near-infrared band). The orientation of the disk leads
to a difference in the time of the center of the disk-only event with 
respect to the star-only event. This time difference, together with the 
impact parameter inferred from the star-only light curve, gives the 
orientation. Once the orientation is known, the duration time, which
corresponds to the chord length, and the magnification, which is related 
to the area of the projected disk, can be used to infer the size and the 
inclination of the disk. For a realistic, non-uniform disk, the surface 
brightness profile has the effect to make the light curve more asymmetric 
if the orientation is neither $0^\circ$ nor $90^\circ$. This also helps
to break the degeneracy between the orientation and the inclination (with
the assumption of a circular, azimuthally symmetric disk in projection).

\subsection{The $\sim$20$\mu$m Transition}

\begin{figure}[h]
\plotone{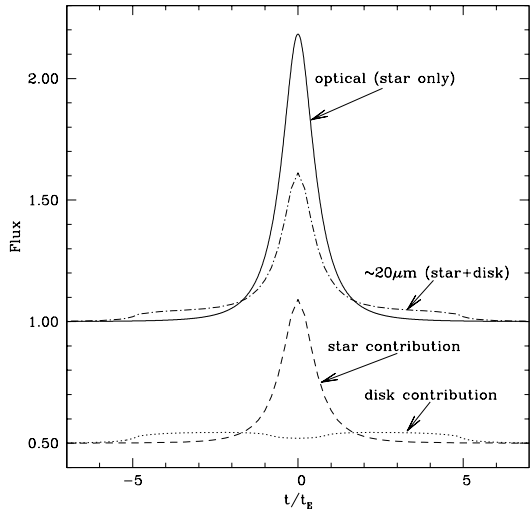}
\caption[]{\label{fig:stardisk}
Contributions of star and disk to the lensing signal. The plot assumes that 
the star and the disk have the same unlensed flux at $\sim 20\mu$m. 
The dot-dashed curve shows the total lensed flux at $\sim 20\mu$m,
which is the sum of that from the star (dashed curve) and that from the 
disk (dotted curve). The solid curve is the lensed flux in optical
where only the star is visible. The lensing of the star+disk system is 
apparently chromatic because the system has different spatial structures
in optical and in mid-infrared.
}
\end{figure}

For a large range of disk-to-star luminosity ratios and temperatures, 
the flux of the star becomes as important as that of the disk in the 
$\sim10-20\mu$m wavelength range. As a point source, the star has its 
flux highly magnified, while the amplification of the disk flux
is, as seen above, at the 10-20\% level and lasts over a longer period.
An interesting phenomenon in this case is that the
lensing event appears to be chromatic if we compare the light curve in
optical and that at $\sim 20\mu$m. The reason is as follows: in
optical, the baseline flux (unlensed flux) is from the star and the
whole flux is magnified by the lens. At $\sim 20\mu$m, the baseline
flux is the sum of those from the star and the disk, but only the
star's flux is highly magnified by the lens.  Therefore, the apparent
magnification of the star+disk system is much less than that for the
star. That is, if the fraction of the unlensed fluxes from the star
and the disk are $x$ and $1-x$, the magnification of the system is
then $xA_\star+(1-x)A_d$, where the magnification $A_\star$ of the star is
much larger than $A_d$, that of the disk.  Figure~\ref{fig:stardisk}
shows an example of the light curve for the system in optical and at
$\sim 20\mu$m with the assumption that the star and the disk have the
same flux ($x=0.5$) at $\sim 20\mu$m. The apparent chromatic effect
can happen if the system to be lensed has different spatial structures
in two bands or there is a color gradient across the (large) system. 

\subsection{Visible and Near-infrared: Reflected Light}

\begin{figure}
\plotone{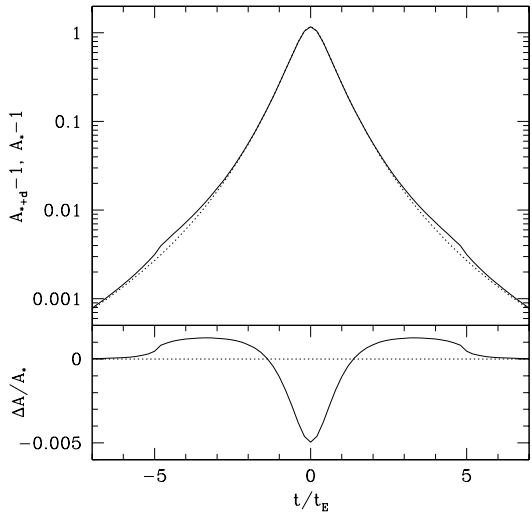}
\caption{The top panel shows the classical microlensing light curve
of a point source (dotted curve) and that of a star plus the disk
in reflected light (solid curve). A fraction of $\tau=10^{-2}$ of the 
star light is assumed to be reflected by the disk.  The bottom panel 
shows the relative flux difference: $\tau\,(A_d/A_\star-1)$. Parameters
of the disk are the same as those used for the solid curve in 
panel (a) of Fig.~\ref{fig:lightcurve}.}
\label{plot_reflected}
\end{figure}

It is interesting to note that constraints on the disk can also be
obtained at shorter wavelengths where the star flux largely dominates:
a fraction $\tau$ of the star light is directly reflected by the 
circumstellar dust (so strictly speaking, the star-only case we mentioned 
before is not exactly so). The reflected light has similar spectral energy 
distribution as that of the star, but follows the spatial distribution of the
dust in the disk. Therefore, the observed light curve carries the
signature of the lensed disk at a small level.  Such a light curve is
illustrated in the top panel of Figure~\ref{plot_reflected}. The
dashed curve shows the classical microlensing light curve of a point
source (the star) and the solid curve shows light curve of the star+disk
system, where a fraction of $\tau=10^{-2}$ of the star's light is seen in
the disk. The light curve shown in Figure~\ref{fig:stardisk} can be thought 
as an extreme version of Figure~\ref{plot_reflected} with $\tau\sim1$. 
The bottom panel of Figure~\ref{plot_reflected} shows 
the relative difference in the magnifications of the star+disk system and 
the star alone, i.e., the quantity $\tau\,(A_d/A_\star-1)$. It is essentially
at the level of $\tau\times 10\%$ near the wings and $\tau\times 100\%$ near
the center of the light curve. The corresponding curve follows similar 
parameter dependences as the ones presented in Figure~\ref{fig:lightcurve1} 
and can be used in the same way to constrain disk properties.

\section{Observational Prospects}

\subsection{Detectability}

In this section we discuss observational requirements for the detection of
microlensing of circumstellar disks. As the ranges of temperature, surface 
brightness, and distance of the systems can be quite large, we will only 
provide some order of magnitude estimates for our fiducial circumstellar 
disk located at 8 kpc from us (see Fig.\ref{fig:sed}) and for a 1$M_\odot$ 
lens located halfway to the Galactic center. Below we will address the 
detectability of the microlensing effects and will consider as an example 
the current specifications of the upcoming James Webb Space Telescope (JWST), 
which is a 6-meter telescope giving access to a wavelength range of 0.6 to 
25.5 $\mu$m.

At the Galactic center, a circumstellar disk with radius as large as
100 AU has an apparent size of about $0.01''$, which is much
smaller than the size of the telescope point spread function (PSF).
As we have discussed, the properties of the disk can be probed with 
microlensing in several ways: from its thermal emission (peaking around 
50$\mu$m) and from the reflected light of the star (peaking around 1$\mu$m). 
We will investigate these two possibilities separately.

The JWST cameras allow to observe in the 0.6-25.5$\mu$m range.
Between 10 and 25$\mu$m, the main noise contributions for observing
distant circumstellar disks originate from the detector self-emission
and the zodiacal light (both are more important than the noise level
of the star). As these two contributions scale linearly with the
exposure time $t_{\rm exp}$, the detection signal-to-noise ratio (S/N)
of the disk is therefore proportional to $\sqrt{t_{\rm exp}}$. In this
wavelength range, it should be noted that the sensitivity of the
detector decreases exponentially as a function of wavelength. Because of 
this dramatic change, systems with warmer disks peaking at shorter 
wavelengths will be much easier to observe.

\begin{figure}
\plotone{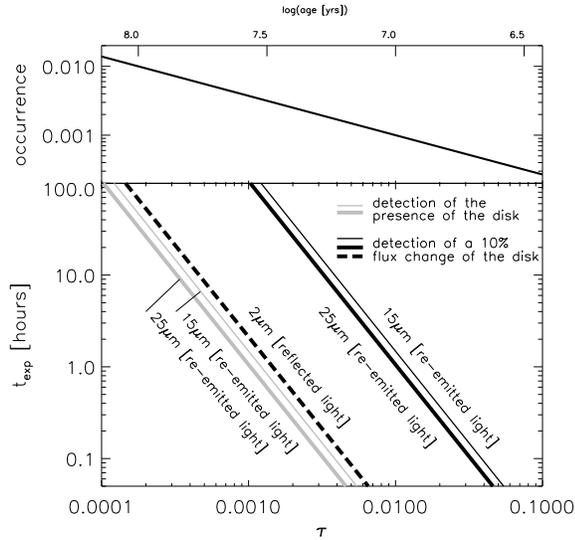}
\caption{The top panel shows, as a function of the disk-to-star luminosity 
ratio $\tau$, the quantity ${\rm age}/10^{10}$yr, corresponding to a rough 
estimate of the occurrence of microlensing events of circumstellar disk 
systems. The bottom panel presents the required exposure time to detect the
disk (gray lines) and to detect a 10\% change in the disk flux (solid black
lines) at the 3$\sigma$ level with the JWST. The thin and thick solid lines
are computed for re-emitted light at 15 and 25.5 $\mu$m respectively
and include noise contributions from the instrument self-emission and
the zodiacal light. The dashed line corresponds to the microlensing
signal seen from the reflected light of the disk at 2 $\mu$m, for
which the main source of noise comes from the star light. }
\label{plot_SN}
\end{figure}

Using the sensitivity and noise contributions given by the JWST Exposure 
Time Calculator for the Mid Infrared Instrument (MIRI) we have estimated 
several quantities: (i) The exposure time required to detect the presence 
of the disk, i.e., to obtain a 3$\sigma$ detection of the infrared excess 
at 15 and 25.5 $\mu$m, respectively. This is shown in the bottom panel of 
Figure \ref{plot_SN} (gray lines) as a function of the disk-to-star 
luminosity ratio $\tau$.  (ii) The exposure time required to have a 
3$\sigma$ detection of a 10\% change in the disk flux, i.e., to detect the 
microlensing signal. This is shown in solid black lines.  
For $10^{-3}<\tau<10^{-2}$, the exposure time ranges between 1 and 100 
hours.  At 10$\mu$m and below, the intensity of the blackbody spectrum
of the disk decreases rapidly and does not allow any detection within a 
reasonable exposure time.  Finally, we also point out that the current 
observational limitations come from instrumental noise. One of the main 
noise contributions of MIRI is the instrument's thermal emission. Contrary 
to the sensitivity in optical bands, that of mid/far-infrared detectors 
is expected to keep improving over the years as a result of the development 
of new technologies. Next generation mid-infrared detectors might therefore
need significantly less exposure time than the ones quoted above
and may even allow us to observe at longer wavelength where the
emission from disk dominates over that of the star.

The second way for using microlensing to extract information about the
disk is to measure the light reflected by the disk. In such a case,
the signal is the slight deviation of the microlensing light curve
from that of a point source, which is typically at the
$\tau\times$10\% level.  We assume that the observation is performed
in the K band ($2.2\mu$m) to minimize the effect of the Galactic
extinction. At this wavelength, the width of the JWST PSF is about 
$0.1''$. A rough estimate based on the Galactic stellar disk profile 
in \citet{zheng2001} shows that, toward the Galactic center, the 
probability that another disk star falls within the PSF is less than 
0.2\%. Inclusion of buldge stars would increase the probability by rougly an order of magnitude.
Therefore, in the case of the lensing of the 
circumstellar disk in reflected light observed by JWST-like space 
telescopes, the main contribution to the noise originates from the light 
of the central star (neglecting instrumental and calibration uncertainties). 
The dashed line in Figure~\ref{plot_SN} shows the exposure time required to 
have a 3$\sigma$ detection of a 10\% change (caused by microlensing) in the 
flux of the reflected light.  For $10^{-3}<\tau<10^{-2}$, the exposure time 
ranges between $2\times 10^{-2}$ and 2 hours.

The duration of the microlensing event for a circumstellar disk is of the 
order of a year. Over such a time scale, a few exposures would allow us to 
sample the disk light curve with enough accuracy to constrain a number of 
parameters of the disk like its inner and outer radii. We note that the 
numbers used in this estimate are based on a typical circumstellar disk 
surface brightness. Brighter disks occur and would therefore reduce the
exposure time mentioned above. Moreover we have based our estimates
on a uniform disk. A steeper surface brightness profile will have a
similar effect as reducing the effective size of the disk and therefore 
increase the amplitude of the magnification.

\subsection{Event Occurrence}

The disk-to-star luminosity ratio $\tau$ of the circumstellar disk is expected 
to decrease as a function of the age of the system. Observationally, it has 
been shown that $\tau \propto {\rm age}^{-1.75}$ \citep{zuckerman2004}. The 
top axis of Figure~\ref{plot_SN} labels the age of a system corresponding to 
the value of $\tau$ on the bottom axis.  We see that the bright phase of the 
circumstellar disk typically lasts for a few to $\sim$100 Myr, therefore 
the fraction of stars that have bright circumstellar disks is determined by 
the fraction of young stars, which is closely related to the star formation 
history.  

For a constant star formation rate over 10 Gyr, the fraction of young
stars is given by the ratio of the stars' age and 10 Gyr. this
fraction can be regarded as an order of magnitude estimate of the
fraction of microlensing events that could be those of star+disk
system, i.e., the occurrence of microlensing events of circumstellar
disk.  This is shown in the top panel of Figure~\ref{plot_SN}.  The
values are only indicative since the star formation history of the
Milky Way is much more complex than a pure continuous star
formation. In the disk, there seems to be 5 episodes of enhanced star
formation over the last 2 Gyr that could be related to density waves
and/or perigalactic passages of satellite galaxies
\citep{marcos2004}. Most young stars stay near the plane where they
were born since they do not have enough time to diffuse
away. Therefore, along lines of sight in the Galactic mid-plane away
from the bulge, the young star fraction is expected to be much higher
than that shown in the top panel of Figure~\ref{plot_SN}.  The inner
part of the Galaxy is thought to be dominated by old bulge stars.
This may be true in the outer bulge. However, H$_2$ clouds and star
forming regions are present (\citealt{bania77,liszt78}) in the
Galactic center. In the central few hundred parsecs of the Galaxy,
there are several lines of evidence showing recent (e.g., 200 Myr ago)
and ongoing star formation (see \citealt{loon2003} and references
therein).  So we also expect a larger occurrence rate of systems with
circumstellar disks for lines of sight toward the very center of the
Galaxy.  Taking these recent star formation histories into account
could increase the plotted occurrence in Figure~\ref{plot_SN} by
approximately an order of magnitude.  A more detailed study is
necessary for an accurate determination of the fraction of young stars
along different lines of sight.

Presently several thousands of microlensing events have been detected toward 
the Galactic bulge by OGLE (e.g., \citealt{udalski2000,wozniak2001,sumi2005}), 
MOA (e.g., \citealt{sumi2003}), MACHO (e.g., \citealt{thomas2004}), and EROS 
(e.g., \citealt{afonso2003b}).  
MOA\footnote{http://www.massey.ac.nz/\symbol{126}iabond/alert/alert.html}, 
OGLE-III\footnote{
http://www.astrouw.edu.pl/\symbol{126}ogle/ogle3/ews/ews.html} 
\citep{udalski2003}, and future experiment will provide us with a large 
amount of data to select 
candidate events of microlensing of circumstellar disks to be followed up in 
near- and mid-infrared. The fields monitored by current microlensing projects 
are along the lines of sight of low extinction regions (e.g., Baade's window), 
which are typically a few degrees away from the Galactic center. To enhance 
the fraction of microlensing events of young stars, surveys toward the very 
center of the Galaxy are desired because both the Galactic center and the 
mid-plane of the disk are expected to have higher fraction of young stars. 
To minimize the extinction, such surveys needs to be done in the K band, as 
advocated by \citet{gould1995}. In addition to the fraction of young stars, 
the microlensing optical depths is also enhanced in these surveys. For 
example, the microlensing optical depth of the Galactic disk is about 3 
times larger toward the Galactic center than toward Baade's window 
(\citealt{gould1995}). With the improvements of infrared detectors (such as 
mosaics of infrared arrays, e.g., \citealt{finger2003,dorn2004,mclean2004}), 
ground-based infrared monitoring programs toward the very center of the 
Galaxy will become possible in the near future. Space-based infrared monitoring 
programs might also start in a near future with for example the Microlensing Planet Finder (MPF, see \citealt{Bennett04}).
They will greatly improve the search for microlensing of circumstellar disks.

Given the different timescales involved in the star and disk lensing, 
it is possible to observe the magnification of the disk several months 
or even a year after the optical microlensing event of the star.
In order to optimize the use of telescope time, the following
observational strategy should be followed: (i) detecting a
microlensing event in the optical or infrared; (ii) if the star is 
likely to reside in a star-forming region or an H$_2$ cloud that indicates a
young age, then following up in the near-infrared (mid-infrared) in order 
to detect the presences of a circumstellar disk in reflected (re-emitted) 
light; (iii) if a disk is present, then performing longer exposures to
sparsely (e.g., once a month) sample the light curve and obtaining 
constraints on the disk properties. In these three steps, the second one is 
crucial. We want the rate of successful selection to be as high as possible.
For this purpose, we can check the (projected) location of the source star 
to see whether it coincides with that of an H$_2$ clouds. We can also check 
the infrared color-magnitude diagram of neighbor stars of the source star 
to see whether there is any hint of young stars (e.g., the existence of 
evolved massive stars). (iv) In addition, one can increase the probability 
of observing a microlensing event of a disk by preselecting events caused by 
massive and nearby lenses (e.g., by choosing long duration microlensing
events) in order to have larger projected Einstein ring radii and hence
higher disk magnifications.

\section{Summary and Discussion}

We have investigated the microlensing effects on a source star surrounded by
a circumstellar disk, as a function of wavelength:
\begin{itemize}

\item In the mid- and far-infrared, where the emission of the system
is dominated by the cold dusty disk, the microlensing light curve
encodes the geometry and surface brightness profile of the disk.  For
a system located at the Galactic center, typical disk magnifications
reach $\sim1.1-1.2$ and the events last over $\sim 1$ year, i.e., much
longer than that of the star-only case.

\item At around 20$\mu$m, where the emission for the star and the disk 
are in general comparable, the difference in the emission areas results 
in a chromatic microlensing event.

\item Finally, in the near-infrared and visible, where the emission of
the star dominates, the small fraction of star light directly
reflected by the disk causes the light curve of the system to slightly 
deviate from that of a point source. In this case, typical departures from 
the classical microlensing light curve of a point source are at a level 
of $\tau\times10\%$, where $\tau$ is the disk-to-star luminosity ratio. 
For young disks, a photometric accuracy of 10$^{-3}$ will allow us
to probe the disk properties.

\end{itemize}

We have shown how the disk light curves depend on a number of disk
parameters like the size, the orientation, the inclination, the radius
of the inner hole, etc., and have pointed out that in some
configurations the magnification can be substantially higher. The most
favorable case is the lensing of a small, highly inclined disk with a
steep surface brightness profile.

Among the detected microlensing events, there is a significant fraction 
of events with binary lens
(e.g., \citealt{alcock2000,jaroszynski2002, jaroszynski2004}).
In the case of a binary lens, if the separation of the binary is much
smaller than the circumstellar disk, the light curve would largely
remain the same as that of a single lens. In addition it will better
resolve the fine features of the disk such as gaps and hot spots, if
they happen to come close to the caustics of the binary lens.

The duality between the infrared and optical regimes and the long
timescale relevant for the disk microlensing allow us to maximize the
observation efficiency by only following up systems that have already
experienced a substantial magnification in the optical. The fraction
of stars exhibiting a bright circumstellar disk that can be detected 
through the lensing effect is expected to be of order a few times 
10$^{-3}$ with large variations among different lines of sight. A
K band micro-lensing survey toward the very center of the Galaxy is 
desired for enhancements in both microlensing optical depth and occurrence 
rate of lensing events of circumstellar disks.

We have investigated the detectability of such events with the
upcoming JWST telescope. For young disks, a few hours of exposure 
every month would allow us to sample the microlensing light curve 
of the disk and extract some disk parameters.  In addition, the 
relatively deep exposure of the fields resulting from these observations 
can be used, as a byproduct, for various studies of objects toward the
Galactic center.

Microlensing enables us to extend the observation of circumstellar
disks to a distance far beyond the solar neighborhood, in regions
of different environments and metallicities. It has the potential
to reveal the diversity of circumstellar disks. Studying the amount and 
nature of their dust and their dependence on the environment would improve
our understanding of the formation and evolution of the circumstellar
disks and planets.

\acknowledgments 
We thank Scott Gaudi, Andy Gould, and Bohdan Paczy\'{n}ski for stimulating 
discussions and useful suggestions. We are also grateful to Roman Rafikov, 
Dejan Vinkovic, Doron Chelouche, James Gunn, and John Bahcall for helpful 
discussions and comments. Z.Z. acknowledges the support of NASA through 
Hubble Fellowship grant HF-01181.01-A awarded by the Space Telescope Science 
Institute, which is operated by the Association of Universities for Research 
in Astronomy, Inc., for NASA, under contract NAS 5-26555. B.M. acknowledges 
the Florence Gould foundation for its financial support.
 
{}

\end{document}